\documentclass[9pt,conference,a4paper]{IEEEtran}
\ifCLASSINFOpdf
   \usepackage[pdftex]{graphicx}
\else
\fi
\usepackage{url}

\usepackage{color}

\begin{document}

\title{Flaglets: Exact Wavelets on the Ball}

\author{%
\IEEEauthorblockN{
Boris Leistedt\IEEEauthorrefmark{1} and  
Jason D. McEwen\IEEEauthorrefmark{1}}
\IEEEauthorblockA{\IEEEauthorrefmark{1} 
Department of Physics and Astronomy, University College London, London WC1E 6BT, U.K.\\
\tt\footnotesize \{boris.leistedt.11, jason.mcewen\}@ucl.ac.uk}
}

\maketitle

\begin{abstract}
We summarise the construction of exact axisymmetric scale-discretised wavelets on the sphere and on the ball. The wavelet transform on the ball
relies on a novel 3D harmonic transform called the Fourier-Laguerre transform which combines the spherical harmonic transform with damped
Laguerre polynomials on the radial half-line. The resulting wavelets, called flaglets, extract scale-dependent, spatially localised features in three-dimensions while treating the tangential and radial structures separately. Both the Fourier-Laguerre and the flaglet transforms are theoretically exact thanks to a novel sampling theorem on the ball. Our implementation of these methods is publicly available \cite{flaglets, s2let2012} and achieves floating-point accuracy when applied to band-limited signals. 
\end{abstract}

\section{Introduction}
Spherical wavelets have been applied successfully to numerous problems in astrophysics and geophysics to extract features of interest from signals on the sphere. But in these disciplines, signals on the sphere are often completed with radial information such as depth, redshift, or distance, in which case a full 3D analysis is required.

\section{Scale-discretised wavelets on the sphere}
Scale-discretised wavelets \cite{s2let2012, mcewen2008dirwavelets} allow one to probe and extract scale-dependent, spatially localised features in signals defined on the sphere. In the axisymmetric case (i.e. azimuthally symmetric when centered on the poles) scale-discretised wavelets reduce to needlets \cite{Marinucci2008needlets} and are constructed through a tiling of the harmonic line, thus defining an exact continuous transform on the sphere. Also, both the spherical harmonic and the scale-discretised wavelet transforms are exact in the discrete setting thanks to the use of a sampling theorem on the sphere \cite{mcewen2011novelsampling}. In other words a band-limited signal, i.e. described by a finite number of spherical harmonic coefficients, is represented by a finite number of samples on the sphere without any loss of information. Since the wavelets are band-limited by construction a multiresolution algorithm is used to speed up the transform by capturing each wavelet scale in the minimal number of samples on the sphere. Our implementation of the scale-dicretised wavelet transform is publicly available in the {\tt S2LET} package \cite{s2let2012} which supports the {\tt C}, {\tt Matlab}, {\tt IDL} and {\tt Java} programming languages. At present the code is restricted to axisymmetric wavelets but will be extended to directional, steerable wavelets and spin functions in a future release. 

\section{Flaglets on the ball}

The starting point to construct scale-discretised wavelets on the ball is a novel 3D transform, the Fourier-Laguerre transform, combining the spherical harmonics with damped Laguerre polynomials on the radial half-line \cite{flaglets}. We construct axisymmetric wavelets, which we call flaglets, by separately tiling the tangential and radial harmonic spaces of the Fourier-Laguerre transform. Both the Fourier-Laguerre and flaglet transforms are exact continuous transforms, which are also exact in the discrete setting thanks to a 3D sampling theorem on the ball. Flaglets extract scale-dependent, spatially localised angular and radial features in signals defined on the ball.  Since the wavelets are band-limit in Fourier-Laguerre space by construction, a multiresolution algorithm is again introduced.  Our implementations of these transforms on the ball are publicly available in the {\tt FLAG} and {\tt FLAGLET} packages \cite{flaglets}.

\section{Applications and perspectives}

\begin{figure}[h]\centering\includegraphics[width=8.9cm]{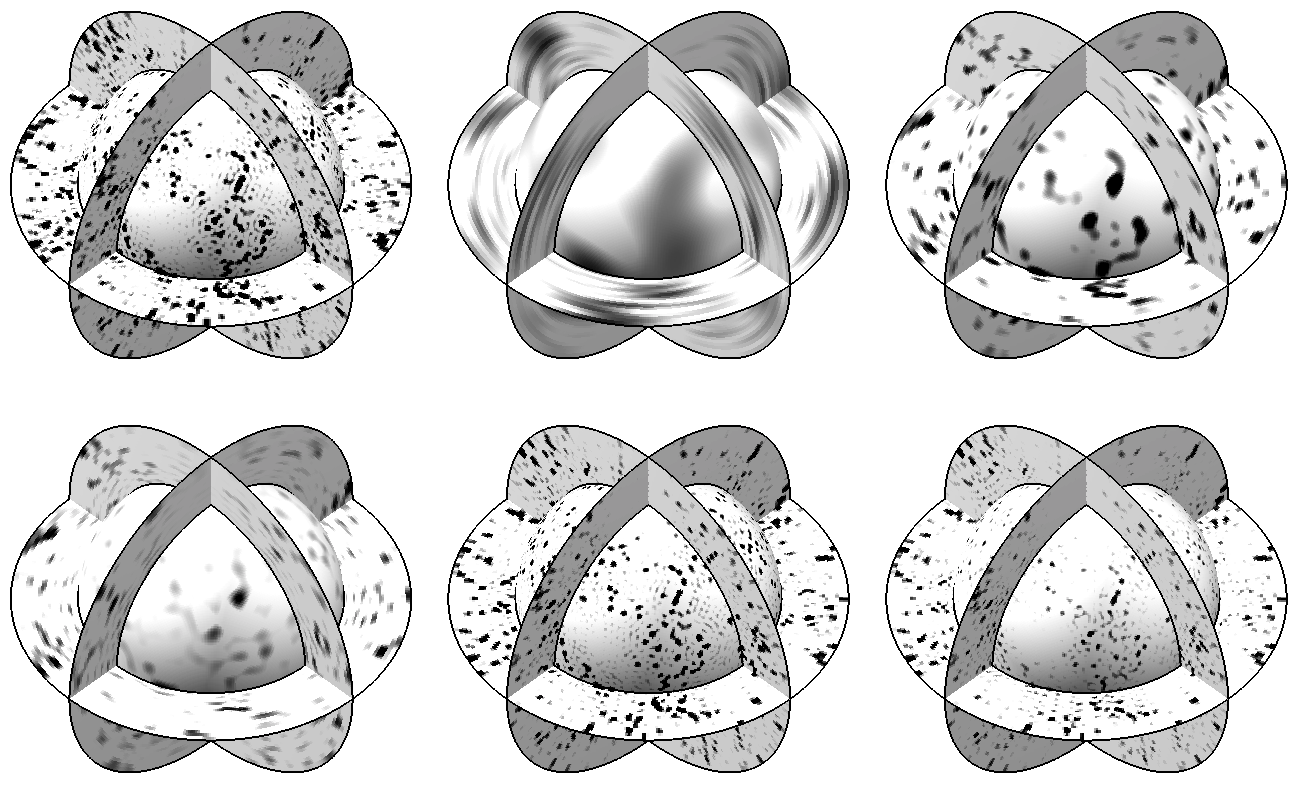}\caption{N-body simulation (top left panel) and its flaglet coefficients for decreasing flaglet scales (subsequent panels from left-to-right, top-to-bottom). }\label{fig:nobody}\end{figure}

The flaglet transform probes tangential and radial structures at scales of interest while capturing all the information of a band-limited signal in the minimal number of samples on the ball. It is suitable for high precision analysis of 3D data that requires the separate treatment of angular and radial components. In future application we intend to exploit flaglets to study galaxy surveys, which are used in cosmology to study the large-scale structure of the Universe, specifically by confronting observations (e.g. clustering properties) with predictions of physical models. Galaxy surveys are contaminated with intrinsic uncertainties and systematics affecting the radial and angular dimensions differently. For example photometric redshifts of galaxies are estimated from colour information with much higher uncertainty than the estimate of the angular position of galaxies. Hence a separate treatment of the angular and radial information is needed to efficiently extract cosmological information from galaxy surveys and to constrain relevant physical theories. Also, the flaglet transform takes advantage of the sparsity of these surveys: gravity tends to generate a filamentary structure that is captured in a small number of flaglet scales, as shown in figure \ref{fig:nobody} for an N-body simulation.

\bibliographystyle{IEEEtran}
\bibliography{bib}

\end{document}